\documentclass[letterpaper, 10 pt, conference]{ieeeconf}  

\IEEEoverridecommandlockouts                              

\usepackage{amsmath,amsfonts,amssymb}
\usepackage{algorithmic}
\usepackage{algorithm}
\usepackage{array}
\usepackage{textcomp}
\usepackage{stfloats}
\usepackage{url}
\usepackage{verbatim}
\usepackage{multirow}
\usepackage{graphicx}
\usepackage{cite}
\graphicspath{{Images/}}
\usepackage{epsfig}

\title{\LARGE \bf
LQ Control of Traffic Flow Models via Variable Speed Limits
}
\author{Brian Block and Stephanie Stockar
\thanks{This material is based upon work supported by the National Science Foundation Graduate Research Fellowship Program under Grant No. DGE-1343012 and NSF CAREER Award 2042354.}
\thanks{The authors are with the Department of Mechanical and Aerospace Engineering and the Center for Automotive Research, The Ohio State University, 930 Kinnear Road, Columbus, OH 43212 USA
        {\tt block.168@osu.edu}}%
}

\begin{document}

\maketitle
\thispagestyle{empty}
\pagestyle{empty}

\begin{abstract}

In this paper, an extension of a linear control design for hyperbolic linear partial differential equations is presented for a first-order traffic flow model. Starting from the Lighthill-Whitham-Richards (LWR) model, variable speed limit control (VSL) is applied through a modification of Greenshield's equilibrium flow model. Then, an optimal linear quadratic (LQ) controller is designed on the linear LWR model. The LQ state feedback function is found via the solution of a Riccati differential equation. Unlike previous studies, the control input is the rate of change of the input, not the input itself. The proposed controller is then verified on both the linear and nonlinear models. In both cases, the controller is able to drive the system to a desired density profile. In the nonlinear application, a higher control gain is needed to achieve similar results as in the linear case.

\end{abstract}

\section{INTRODUCTION}
Congestion is a growing concern in major cities as it leads to increased energy usage and pollution \cite{schrank2012umr}. In 2022, the average commuter in the Unites States spent around 73 extra hours in traffic. This is up over 40\% from 2016 \cite{pishue2022inrix, cookson2017inrix}. The modeling of traffic flow and its control offer opportunities to mitigate congestion, improve travel time, and reduce overall energy usage associated with the transportation sector.

One approach to modeling traffic flow in cities and other large scale systems like highways, is through a macroscopic model that looks at the density and flow of traffic \cite{treiber2012traffic}. A common macroscopic traffic flow model is the Lighthill-Whitham-Richards (LWR) model, which consists of a hyperbolic conservation partial differential equation (PDE) in density \cite{lighthill1955kinematic, richards1956shock}. As a first-order model, the LWR model relies on an equilibrium flow equation to determine the velocity of vehicles \cite{treiber2012traffic}. The equilibrium flow-density relationship can be modeled various ways including using a triangular relationship, Greenshield's model, or an exponential model \cite{tumash2019robust,knoop2013traffic,carlson2011local}. The LWR model is commonly used for large scale traffic congestion control as it is both simple and accurate \cite{spiliopoulou2014macroscopic, yu2020bilateral, liu2021robust, tumash2019robust}.

Macroscopic traffic congestion mitigation strategies rely on either boundary control, such as ramp metering, or in-domain control, such as using variable speed limits (VSL). For example, ramp metering is used in \cite{yu2018adaptive, yu2019traffic} where backstepping control is used to drive a continuous traffic flow model to a density setpoint. For the same purpose, a model free reinforcement learning approach is also used in \cite{yu2019reinforcement}. Similarly, proportional-integral control is used for ramp metering \cite{pasquale2018closed} for a discrete traffic flow model. In addition to tracking a density setpoint, time and space dependent density trajectories are tracked via boundary control in \cite{tumash2019robust, tumash2021boundary} with the goal of removing excess cars from the road and matching desired inflows and outflows.

Alternatively to ramp metering, VSL control relies on changing the speed limits within a stretch of highway. The speed limit can be changed such that the flow entering a traffic jam can be lessened or the flow leaving a traffic jam can be increased so that the jam is mitigated faster. Both ramp metering and VSL control are used in \cite{papamichail2008integrated} to reduce on-ramp queues and total travel time of vehicles within highway links. Model predictive control is also used to reduce total travel time through both ramp metering and VSL control \cite{liu2016model}. Both of these approaches use discrete traffic models. VSL control on a continuous PDE is used in \cite{karafyllis2019feedback} by developing feedback control laws that stabilize a desired density profile.

A possible way to realize VSL control that has not been explored is through a linear quadratic regulator (LQR). In previous traffic applications, LQR has been used in conjunction with non-cooperative Nash games to balance the flow of vehicles \cite{pisarski2015nash}. This application, though, uses boundary feedback and is developed on a discrete model, not a continuous traffic flow model. In other hyperbolic PDE applications, LQR has been used as well. In \cite{moghadam2010lq}, an infinite-dimensional LQ controller is designed for boundary control of a system containing a continuous stirred tank and a plug flow reactor. In \cite{aksikas2009lq}, a state feedback operator that controls the distributed jacket temperature of a reactor is computed by solving a matrix Riccati differential equation. Both applications have the same purpose of achieving a desired chemical concentration. 

Most in-domain VSL control of traffic flow has been studied on discretized models \cite{carlson2011mainstream, carlson2011local, papamichail2008integrated, liu2016model}. The approach in \cite{karafyllis2019feedback} was the first to apply VSL continuously in space and time. While LQR for infinite dimensional systems has been proven effective in other areas, its application to in-domain traffic control poses a challenge because the rate of change of the control input is paramount. The speed limit cannot change too quickly over space as that creates unrealistic driving behavior such as large changes in the acceleration of vehicles. This paper presents a novel LQR formulation where instead of controlling the input, the rate of change is controlled for a continuous hyperbolic PDE model. This approach also differs from previous continuous VSL approaches \cite{karafyllis2019feedback} as, instead of controlling the VSL rate, the rate of change of the speed limit over the length of highway is controlled.


\section{MODEL DESCRIPTION}

\subsection{Lighthill-Whitham-Richards Model}

\begin{figure}[thpb]
  \centering
  \includegraphics[width=0.75\columnwidth]{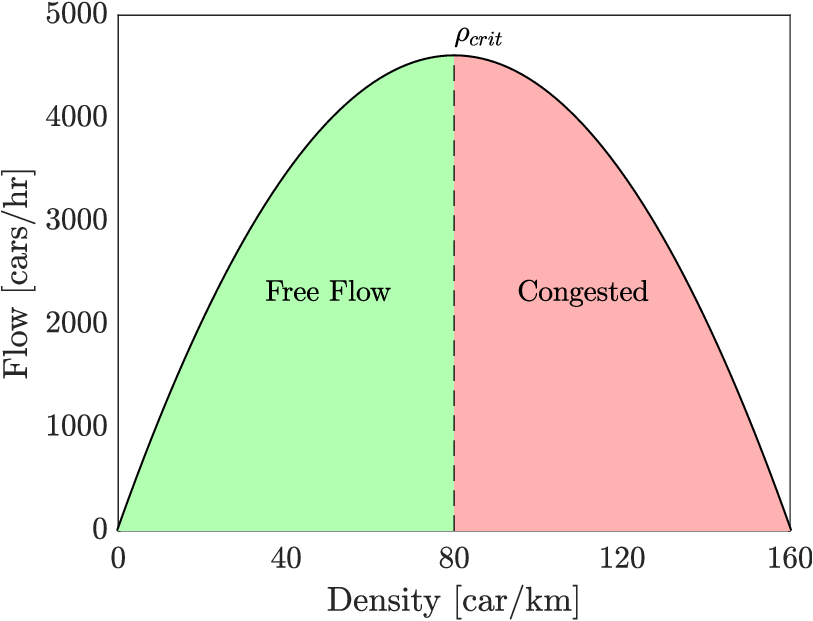}
  \caption{Fundamental diagram showing free flow region (green) and congested region (red).}
  \label{fig:fd}
\end{figure}

The LWR model \cite{lighthill1955kinematic, richards1956shock} is a conservation of mass equation given by
\begin{equation} \label{eq:LWR}
    \frac{\partial\rho}{\partial t} + \frac{\partial q}{\partial z} = 0
\end{equation}
where $\rho(z,t)$ is the traffic density and $q(z,t)$ is a fundamental diagram which gives the equilibrium relationship between traffic density and flow. The fundamental diagram is defined as
\begin{equation} \label{eq:FD}
    q = \rho u
\end{equation}
where $u = U(\rho)$ is the equilibrium relationship between density and traffic velocity. A common equilibrium relationship is Greenshield's model which is a decreasing function of density and is given as
\begin{equation} \label{eq:Greenshield}
    U(\rho) = U_{max}\bigg(1 - \frac{\rho}{\rho_{max}}\bigg)
\end{equation}
where $U_{max}$ is the maximum velocity and $\rho_{max}$ is the maximum density. The relationship between traffic flow and density using Greenshield's model is shown in Fig. \ref{fig:fd}. In Greenshield's model, the critical density, $\rho_{crit}$, which is where traffic goes from free flow to congested, is half of $\rho_{max}$.

\subsection{Linearized LWR Model}

The linearized LWR model is obtained by defining the perturbation variables as
\begin{equation} \label{eq:perturbrho}
    \rho = \rho_0 + \Delta\rho
\end{equation}
\begin{equation} \label{eq:pertubu}
    u = u_0 + \Delta u = U_{max}\bigg(1 - \frac{\rho_0+\Delta\rho}{\rho_{max}}\bigg)
\end{equation}
Then, by substituting \eqref{eq:perturbrho},\eqref{eq:pertubu} into \eqref{eq:LWR} the linear LWR model can be written as
\begin{equation} \label{eq:linLWR1}
    \frac{\partial\Delta\rho}{\partial t} + u_0\frac{\partial\Delta\rho}{\partial z} + \rho_0\frac{\partial\Delta u}{\partial z} = 0 
\end{equation}
Expanding all the terms gives
\begin{multline}
        \frac{\partial\Delta\rho}{\partial t} + U_{max}\bigg(1 - \frac{\rho_0}{\rho_{max}}\bigg)\frac{\partial\Delta\rho}{\partial z} + \\
        \rho_0\frac{\partial}{\partial z}U_{max}\bigg(1 - \frac{\Delta\rho}{\rho_{max}}\bigg) = 0 \label{eq:linLWR2}
\end{multline}
which simplifies to
\begin{equation} \label{eq:linLWR3}
    \frac{\partial\Delta\rho}{\partial t} + U_{max}\bigg(1 - 2\frac{\rho_0}{\rho_{max}}\bigg)\frac{\partial\Delta\rho}{\partial z} = 0
\end{equation}

\section{CONTROL FORMULATION}

\subsection{Variable Speed Limit Control}
Variable speed limit control involves changing the speed limit in order to control the flow of traffic \cite{papamichail2008integrated, carlson2011local, carlson2011mainstream,karafyllis2019feedback}. In this work, VSL is added to Greenshield's model to affect $U_{max}$ and thus $q$. The equilibrium velocity \eqref{eq:Greenshield} then becomes
\begin{equation} \label{eq:VSLGreenshield}
    U_{VSL}(\rho) = b U_{max}\bigg(1 - \frac{\rho}{\rho_{max}}\bigg)
\end{equation}
where $b$ is the VSL rate. The maximum speed $U_{max}$ acts as the speed limit and the VSL rate $b$ works as the  control input to either decrease or increase the maximum speed. The definition of the control input $b$ introduces a new nonlinearity into the model. Following the same procedure as before, perturbed quantities for the rate of change of the speed limit and velocity are introduced
\begin{equation} \label{eq:perturbb}
    b = b_0 + \Delta b
\end{equation}
\begin{equation} \label{eq:perturbu_vsl}
    u = (b_0 + \Delta b)U_{max}\bigg(1 - \frac{\rho_0 + \Delta\rho}{\rho_{max}}\bigg)
\end{equation}
The perturbation of density stays unchanged from \eqref{eq:perturbrho}. So, \eqref{eq:linLWR2} now becomes
\begin{equation}  \label{eq:linLWR1_vsl}
    \frac{\partial\Delta\rho}{\partial t} + b_0 U_{max}\bigg(1 - \frac{\rho_0}{\rho_{max}}\bigg)\frac{\partial\Delta\rho}{\partial z} + \rho_0\frac{\partial\Delta u}{\partial z} = 0
\end{equation}
Inserting \eqref{eq:perturbu_vsl} into \eqref{eq:linLWR1_vsl} gives
\begin{multline} \label{eq:linLWR2_vsl}
    \frac{\partial\Delta\rho}{\partial t} + b_0 U_{max}\bigg(1 - \frac{\rho_0}{\rho_{max}}\bigg)\frac{\partial\Delta\rho}{\partial z} + \\
    \rho_0 U_{max}\bigg(1 - \frac{\rho_0}{\rho_{max}}\bigg)\frac{\partial \Delta b}{\partial z} - b_0 U_{max}\frac{\rho_0}{\rho_{max}}\frac{\partial\Delta\rho}{\partial z} = 0
\end{multline}
which simplifies to
\begin{multline} \label{eq:linLWR3_vsl}
        \frac{\partial\Delta\rho}{\partial t} + b_0 U_{max}\bigg(1 - 2\frac{\rho_0}{\rho_{max}}\bigg)\frac{\partial\Delta\rho}{\partial z} + \\
        \rho_0 U_{max}\bigg(1 - \frac{\rho_0}{\rho_{max}}\bigg)\frac{\partial \Delta b}{\partial z} = 0
\end{multline}

The variable $\frac{\partial\Delta b}{\partial z}$ is the change in the perturbation of the VSL rate over the length of road. When there is no VSL, $\frac{\partial\Delta b}{\partial z} = 0$ and $b_0 = 1$, and \eqref{eq:linLWR3_vsl} returns to \eqref{eq:linLWR3}.

\subsection{Optimal Control Design for Hyperbolic PDE model}

The first step to derive the state feedback control policy is to write the PDE in its equivalent state space formulation on a Hilbert space $\mathcal{H}$ \cite{aksikas2009lq}. For a linear hyperbolic PDE of the form
\begin{align} \label{eq:hyperPDE}
    \frac{\partial x}{\partial t} &= V\frac{\partial x}{\partial z} + Mx + B_0u \\
    y &= C_0x
\end{align}
this results in
\begin{align} \label{eq:pdeSS}
    \dot{x}(t) &= Ax(t) + Bu(t) \\
    y(t) &= Cx(t)
\end{align}
where $A$ is the linear operator defined as
\begin{equation} \label{eq:A_def}
    A = V\cdot\frac{d.}{dz} + M\cdot I
\end{equation}
on the domain
\begin{equation} \label{eq:domain}
    D(A) = \{x\in\mathcal{H}:\frac{dx}{dz}\in\mathcal{H}\textrm{ and }x(0)=0\}
\end{equation}
with $V\in\mathbb{R}^{n\times n}$ being a symmetric matrix and $M$, $B_0$, and $C_0$ being real continuous space-varying matrix functions. In addition, $x$ is assumed to be absolutely continuous. It is proven in \cite{aksikas2009lq} that if $V<0$ then $A$ generates an exponentially stable $C_0$-semigroup. 

The linear-quadratic optimal control problem on an infinite time interval finds a control input $u_{opt}$ such that the functional $J(x_0,u)$ is minimized. The cost functional is given as
\begin{equation} \label{eq:costfunct}
    J(x_0,u) = \int_{0}^{\infty} \langle Cx(t),QCx(t) \rangle + \langle u(t), Ru(t) \rangle \,dt
\end{equation}
The solution of \eqref{eq:costfunct} can be found by finding the non-negative self-adjoint operator $P$ which solves
\begin{equation} \label{eq:ORE}
    [A^*P + PA + C^*QC - PBR^{-1}B^*P]x = 0
\end{equation}
When $(A,B)$ is exponentially stabilizable and $(Q^{\frac{1}{2}}C,A)$ is exponentially detectable \eqref{eq:ORE} has a unique, non-negative solution $P$ and \eqref{eq:costfunct} is minimized by the unique control input $u_{opt}$ given by
\begin{equation} \label{eq:u_opt1}
    u_{opt}(t)=K_0x(t)
\end{equation}
where the optimal feedback is
\begin{equation} \label{eq:optfeedback}
    K_0=-R^{-1}B^*P
\end{equation}
The proof that $(A,B)$ is exponentially stabilizable and $(P^{\frac{1}{2}}C,A)$ is exponentially detectable given $V<0$ is shown in \cite{aksikas2009lq}. Furthermore,
\begin{equation} \label{eq:Q_0}
    P:=\Phi(z)I
\end{equation}
where $\Phi(z)$ is the solution of the matrix Riccati differential equation given by
\begin{align} \label{eq:MRDE}
    \begin{split}
        V\frac{d\Phi}{dz} &= M^*\Phi + \Phi M + C_0^*Q_0C_0 - \Phi B_0R_0^{-1}B_0^*\Phi\\
        \Phi(L) &= 0
    \end{split}
\end{align}
where $Q_0$ is a positive matrix and $R_0$ is a self-adjoint positive matrix. Thus, \eqref{eq:optfeedback} becomes
\begin{equation} \label{eq:k_0formula}
    K_0 = -R_0^{-1}(z)B^*_0(z)\Phi(z)I
\end{equation}

\subsection{LQ Control Deisgn for LWR Model with VSL}

In order to find $\Phi(z)$, we take \eqref{eq:linLWR3_vsl} and transform it into
\begin{multline} \label{eq:linLWR4_vsl}
        \frac{\partial\Delta\rho}{\partial t} =  -b_0 U_{max}\bigg(1 - 2\frac{\rho_0}{\rho_{max}}\bigg)\frac{\partial\Delta\rho}{\partial z} - \\
        \rho_0 U_{max}\bigg(1 - \frac{\rho_0}{\rho_{max}}\bigg)\frac{\partial \Delta b}{\partial z}
\end{multline}
so that it is now in the form of \eqref{eq:hyperPDE}. It follows that
\begin{align} \label{eq:LWRLQRvariables}
    \begin{split}
        V &= - b_0 U_{max}\bigg(1 - 2\frac{\rho_0}{\rho_{max}}\bigg)\\
        M &= 0\\
        B_0 &= -\rho_0 U_{max}\bigg(1 - \frac{\rho_0}{\rho_{max}}\bigg)\\
        C_0 &= I
    \end{split}
\end{align}
The condition of $V<0$ is met if $\rho_0<\frac{1}{2}\rho_{max}$ in \eqref{eq:LWRLQRvariables}. This means that $\rho_0$ must always fall in the free flow regime, which is a valid assumption since the equilibrium profile should never fall in the congested regime to avoid traffic jams. The control input for the LWR model with VSL is the change of the speed limit over the stretch of highway. Using \eqref{eq:MRDE} and \eqref{eq:LWRLQRvariables} the matrix Riccati equation becomes
\begin{align} \label{eq:riccatiLWR}
    \begin{split}
        V\frac{d\Phi}{dz} &= Q_0 - \Phi B_0R_0^{-1}B_0^*\Phi\\
        \Phi(L) &= 0
    \end{split}
\end{align}
As in \cite{aksikas2009lq}, $R_0=1$. The values for $V$ and $B_0$ will be inserted into the equation at the end of the derivation for $\Phi$ for simplicity. Using separation of variables, \eqref{eq:riccatiLWR} becomes
\begin{equation} \label{eq:lwr_int}
    \int_{0}^{\Phi} \frac{V}{Q_0-B_0^2\Phi^2}\,d\Phi = \int_{0}^{z}\,dz = z + C
\end{equation}
The left hand side of \eqref{eq:lwr_int} can be solved by a combination of integration by partial fractions and substitution. After applying both methods, \eqref{eq:lwr_int} becomes
\begin{equation} \label{eq:solving_for_phi}
    \frac{V}{2B_0\sqrt{Q_0}}(\ln|B_0\Phi+\sqrt{Q_0}|-\ln|B_0\Phi-\sqrt{Q_0}|)=z+C
\end{equation}
Then,
\begin{align} \label{eq:moresolvingforphi}
    \begin{split}
        \ln\bigg|\frac{B_0\Phi+\sqrt{Q_0}}{B_0\Phi-\sqrt{Q_0}}\bigg| &= \frac{2B_0\sqrt{Q_0}}{V}(z+C)\\
        \bigg|\frac{B_0\Phi+\sqrt{Q_0}}{B_0\Phi-\sqrt{Q_0}}\bigg| &= \exp\bigg(\frac{2B_0\sqrt{Q_0}}{V}(z+C)\bigg)\\
        \big|B_0\Phi\!+\!\sqrt{Q_0}\big|\! &=\! \big|B_0\Phi\!-\!\sqrt{Q_0}\big|\exp\bigg(\frac{2B_0\sqrt{Q_0}}{V}(z\!+\!C)\bigg)
    \end{split}
\end{align}
The solution to \eqref{eq:moresolvingforphi} can be either
\begin{equation} \label{eq:incorrectphi}
    B_0\Phi+\sqrt{Q_0} = (B_0\Phi-\sqrt{Q_0})\exp\bigg(\frac{2B_0\sqrt{Q_0}}{V}(z+C)\bigg)
\end{equation}
or
\begin{equation} \label{eq:correctphi}
    B_0\Phi+\sqrt{Q_0} = -(B_0\Phi-\sqrt{Q_0})\exp\bigg(\frac{2B_0\sqrt{Q_0}}{V}(z+C)\bigg)
\end{equation}
Because of the condition that $\Phi(L)=0$, only \eqref{eq:correctphi} can be the solution for $\Phi$. The final solution for the state feedback function $\Phi$ using the condition $\Phi(L)=0$ is
\begin{equation} \label{eq:finalsolutionphi}
    \Phi = \frac{\sqrt{Q_0}\bigg(\exp(\frac{2B_0\sqrt{Q_0}}{V}(z-L))-1\bigg)}{B_0\bigg(\exp(\frac{2B_0\sqrt{Q_0}}{V}(z-L))+1\bigg)}
\end{equation}
Combining \eqref{eq:u_opt1}, \eqref{eq:k_0formula}, and \eqref{eq:finalsolutionphi} with $x(t)=\Delta\rho$ for the linear model leaves us with the final solution for the optimal control input
\begin{equation} \label{eq:u_opt_final}
    u_{opt}(z,t)\! =\! \frac{\partial b}{\partial z}\! =\! -\sqrt{Q_0}\frac{\bigg(\exp(\frac{2B_0\sqrt{Q_0}}{V}(z-L))-1\bigg)}{\bigg(\exp(\frac{2B_0\sqrt{Q_0}}{V}(z-L))+1\bigg)}\Delta\rho
\end{equation}

\begin{figure}[thpb]
  \centering
  \includegraphics[width=0.75\columnwidth]{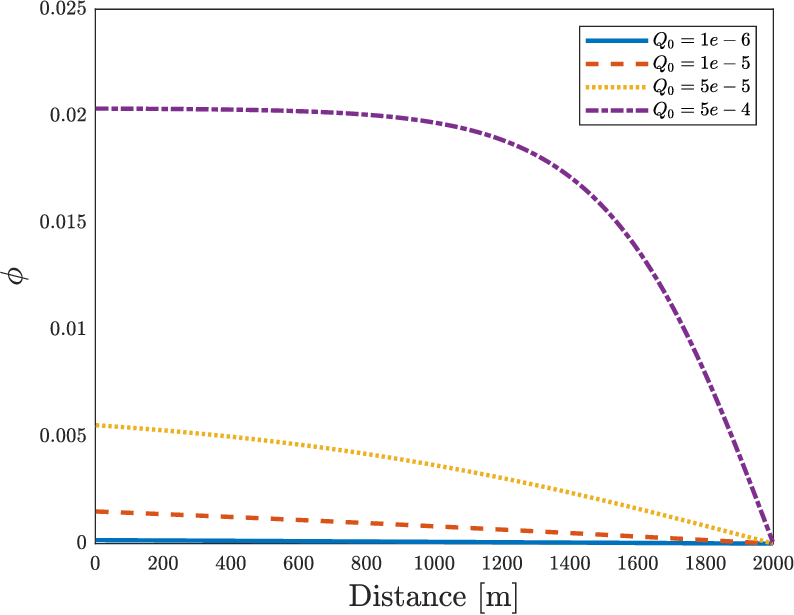}
  \caption{Influence of $Q_0$ on the state feedback function, $\Phi$.}
  \label{fig:phi}
\end{figure}

The parameter $Q_0$ is left as a tuning parameter and its influence on the solution of $\Phi$ is shown in Fig. \ref{fig:phi}.

\section{SIMULATION RESULTS}
\subsection{Uncontrolled Scenario}
The case study investigated in this paper simulates a stretch of highway with an oscillating and increasing flow entering it. This situation could occur because of both on-ramp flow and traffic light operation. Because the base model used for the LQ control is linear, the simulation is kept within the free flow region so that no traffic jams, or shocks in the system, occur. The initial density on the stretch of highway is
\begin{equation} \label{eq:IC}
    \rho(0,z) = 10\sin\bigg(\frac{\pi z}{L}\bigg) + \rho_0
\end{equation}
and the upstream boundary condition is
\begin{equation} \label{eq:BC}
    \rho(t,0) = 5\exp(-L\cdot10^{-6}t)\sin\bigg(\frac{\pi t}{20}\bigg) + \rho_0 + \frac{t}{4L}
\end{equation}

\begin{figure} [ht]
  \centering
  \begin{tabular}{c c}
    \includegraphics[width=.45\columnwidth]{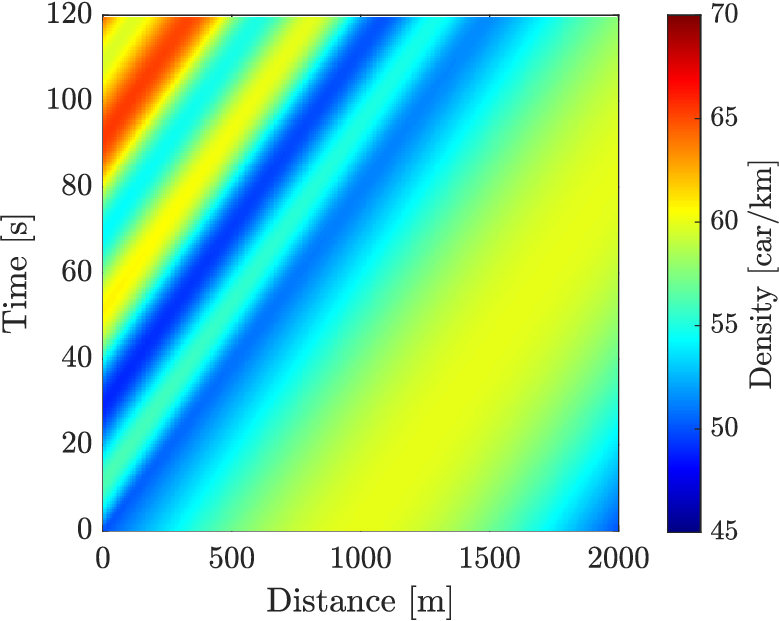} &
    \includegraphics[width=.45\columnwidth]{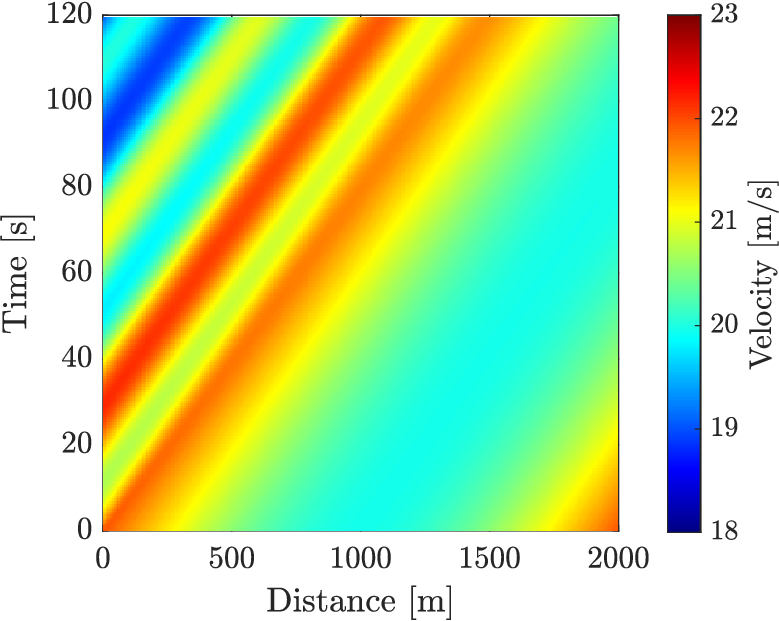}\\
    \small (a) &
    \small (b) \\
    \includegraphics[width=.45\columnwidth]{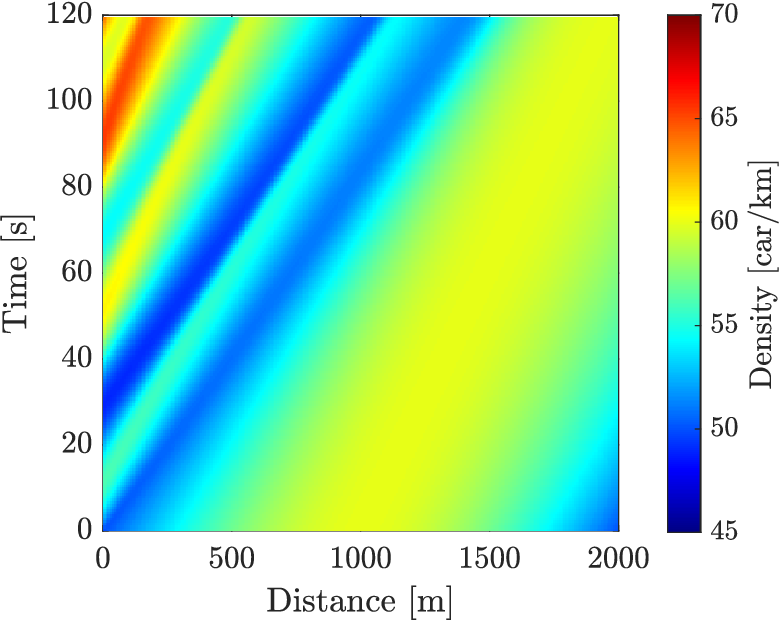} &
    \includegraphics[width=.45\columnwidth]{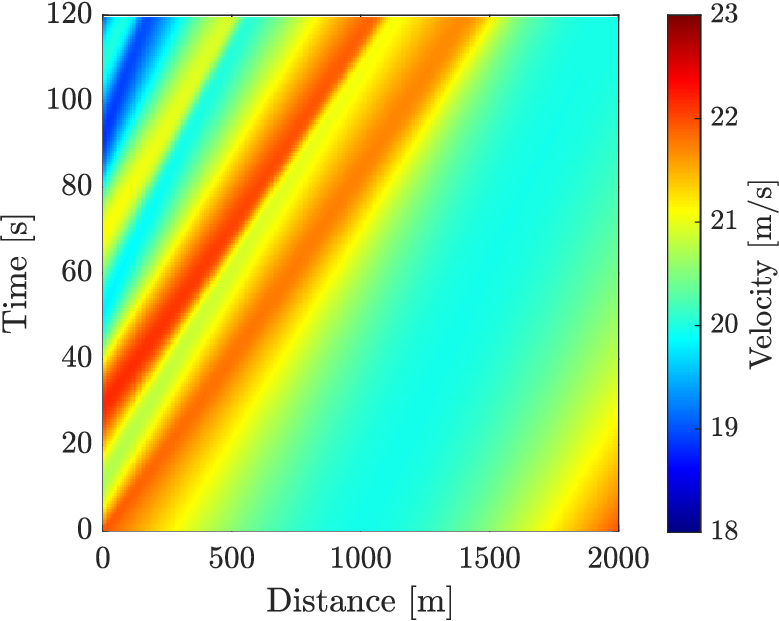}\\
    \small (c) &
    \small (d) 
  \end{tabular}
  \caption{Baseline results for linear model (a), (b) and nonlinear model (c), (d) given the initial and boundary conditions in \eqref{eq:IC}, \eqref{eq:BC}.}
  \label{fig:baseline}
\end{figure}

The model parameters for the simulation are given in Table \ref{tab:params}. Based off of the fundamental diagram shown in Fig. \ref{fig:fd} and parameters in Table \ref{tab:params}, this means that the the density across the highway must be kept below $\rho_{crit} = 80$ cars/km. The resulting density evolution and velocity of the highway section is shown in Fig. \ref{fig:baseline}.
\begin{table}[!ht]
    \centering
    \begin{center}
    \caption{Model Parameters.}
    \label{tab:params}
    \begin{tabular}{l|l|l }
    \hline \hline
    Parameter & Value & Unit\\  \hline 
    Maximum density $\rho_{max}$ & 160 & [cars/km] \\
    Maximum speed $U_{max}$ & 115 & [kph] \\
    Average density $\rho_0$ & 50 & [cars/km] \\
    Road length $L$ & 2000 & [m]\\
    Simulation time $T$ & 120 & [s]\\
     \hline \hline
    \end{tabular}
    \end{center}
\end{table}

The results of the baseline simulation using the linear model are shown in Fig. \ref{fig:baseline}(a),(b), and the results of the baseline nonlinear simulation are shown in Fig. \ref{fig:baseline}(c),(d). For both models, the baseline scenario results stay within the free flow regime as the maximum density reached is less than $\rho_{crit}$. There is good agreement between the results of the linear and nonlinear model. As the density at the beginning of the length of road approaches $\rho_{crit}$ at around 100 s, the solutions between the two models start to differ.

\subsection{Linear LQR Results}
\begin{figure}[thpb]
  \centering
  \includegraphics[width=0.75\columnwidth]{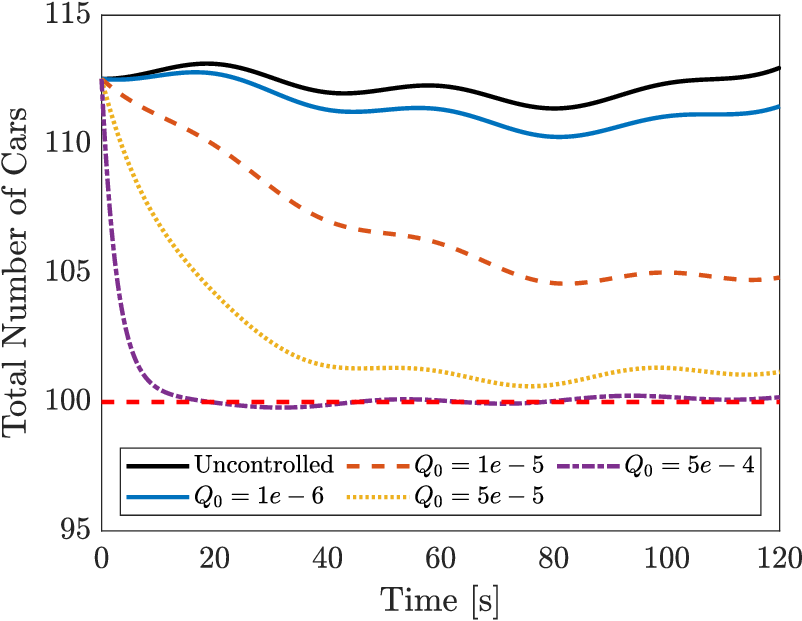}
  \caption{Total cars on highway section for varying values of $Q_0$ when control is applied to linear model.}
  \label{fig:lin_totcars}
\end{figure}

The performance of the LQ controller is first investigated using the linear LWR model. In order to evaluate the performance, the total number of cars in the highway section is calculated as
\begin{equation} \label{eq:totcars}
    \#\textrm{ of cars}= \int_{0}^{L} \rho\,dz
\end{equation}
As the average desired density is 50 cars/km, the desired amount of cars left in the highway section is 100 cars. The results of applying the developed LQ controller to the linear model are shown in Fig. \ref{fig:lin_totcars}. Four different values of the tuning parameter $Q_0$ are investigated. For $Q_0 = 1e-6$, the control is not enough to lower the density by a reasonable amount. Towards the end of the simulation, at around 100 s, the density even starts to rise. For both $Q_0 = 1e-5$ and $Q_0= 5e-5$ the control is enough to lower the density, but at a certain point it no longer drives the density to the desired amount of cars left on the road. For $Q_0 = 1e-5$ this stagnation happens at 80 s and for $Q_0 = 5e-5$ this happens at around 40 s. One thing to note, though, is that when $Q_0 = 5e-5$ the amount of cars does approach the desired amount, but never reaches it. The controller is able drive the linear system to the desired density when $Q_0 = 5e-4$, and it reaches it at around 20 s.

\subsection{Nonlinear LQR Results}
\begin{figure}[thpb]
  \centering
  \includegraphics[width=0.75\columnwidth]{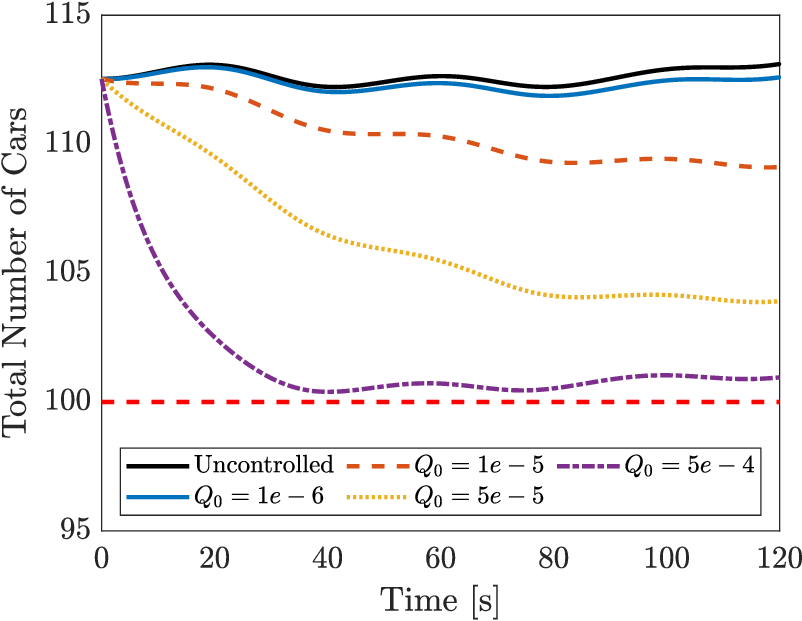}
  \caption{Total cars on highway section for varying values of $Q_0$ when control is applied to nonlinear model.}
  \label{fig:nonlin_totcars}
\end{figure}

For the nonlinear model, the control input is no longer $\frac{\partial b}{\partial z}$, instead it is the actual VSL rate, $b$. So, in order to apply the control to the nonlinear model, first the linear model is used to determine the optimal $\frac{\partial b}{\partial z}$, then the nonlinear control is found by
\begin{equation} \label{eq:nonlincontrolinput}
    b_{opt} = \int_{0}^{L} \frac{\partial b}{\partial z}\,dz
\end{equation}
The nonlinear model \eqref{eq:LWR} then uses \eqref{eq:VSLGreenshield} for the equilibrium velocity equation where $b$ is defined as in \eqref{eq:nonlincontrolinput}. The results for the same values of $Q_0$ as used before, but this time on the nonlinear model, are shown in Fig. \ref{fig:nonlin_totcars}. For $Q_0 = 1e-6$, the system is mostly unaffected. When $Q_0 = 1e-5$ and $Q_0 = 5e-5$ the traffic is reduced, but it does not reach the desired density. With $Q_0 = 5e-4$, the controller is able to get close to the desired density by 40 s, but does not achieve it.

\begin{figure*} [t]
  \centering
  \begin{tabular}{c c c}
    \includegraphics[width=.6\columnwidth]{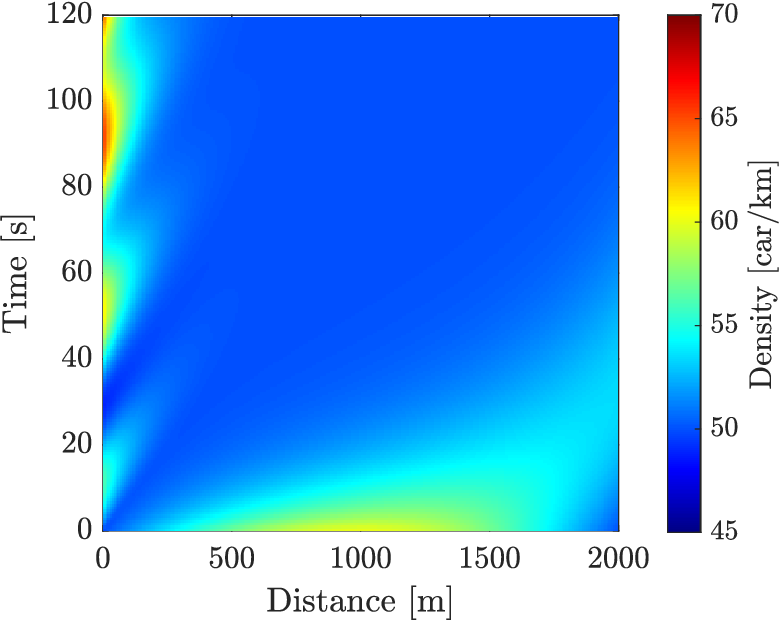} &
    \includegraphics[width=.6\columnwidth]{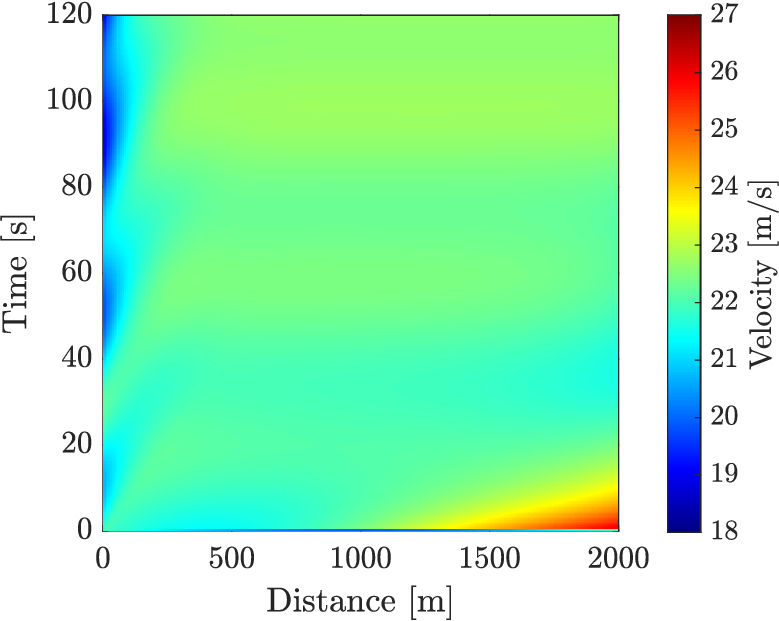} &
    \includegraphics[width=.6\columnwidth]{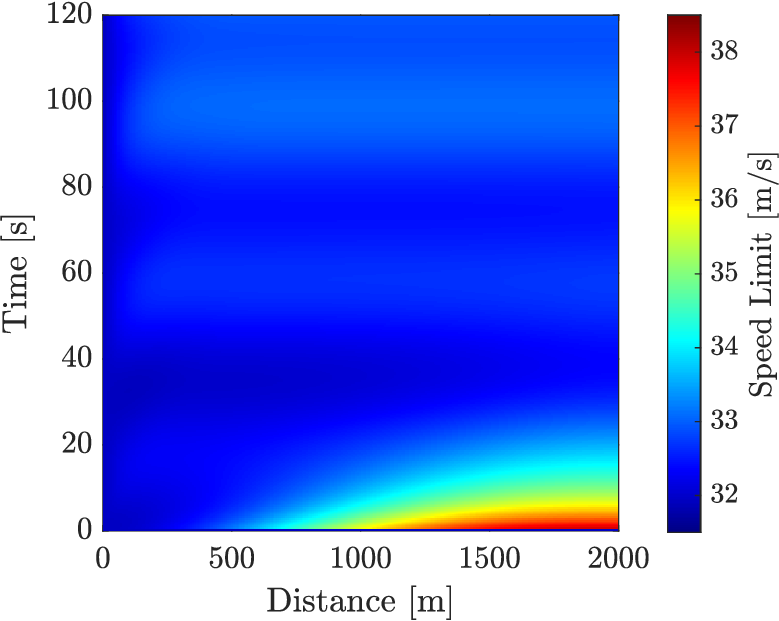} \\
    \small (a) &
    \small (b) &
    \small (c)
  \end{tabular}
  \caption{Density (a), velocity (b), and variable speed limit (c) of linear model with $Q_0 = 5e-5$.}
  \label{fig:lin_cont_results}
\end{figure*}
\begin{figure*} [t]
  \centering
  \begin{tabular}{c c c}
    \includegraphics[width=.6\columnwidth]{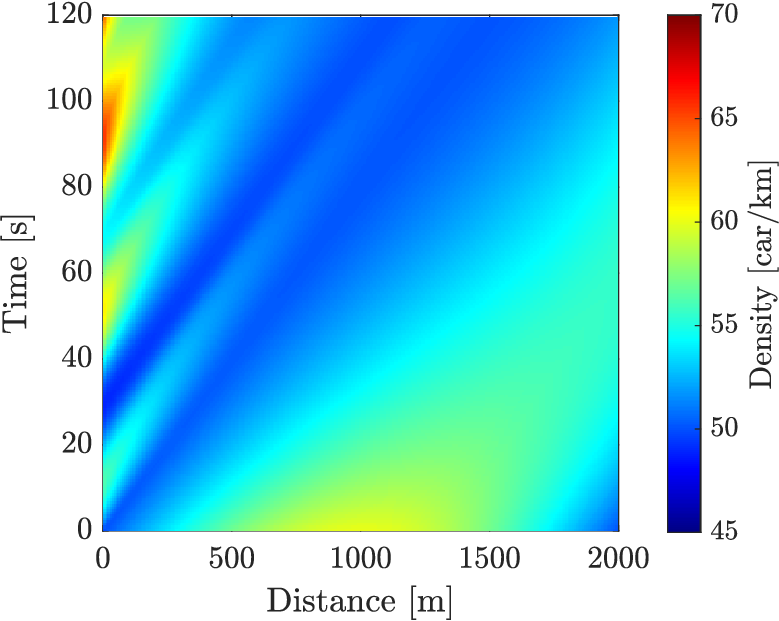} &
    \includegraphics[width=.6\columnwidth]{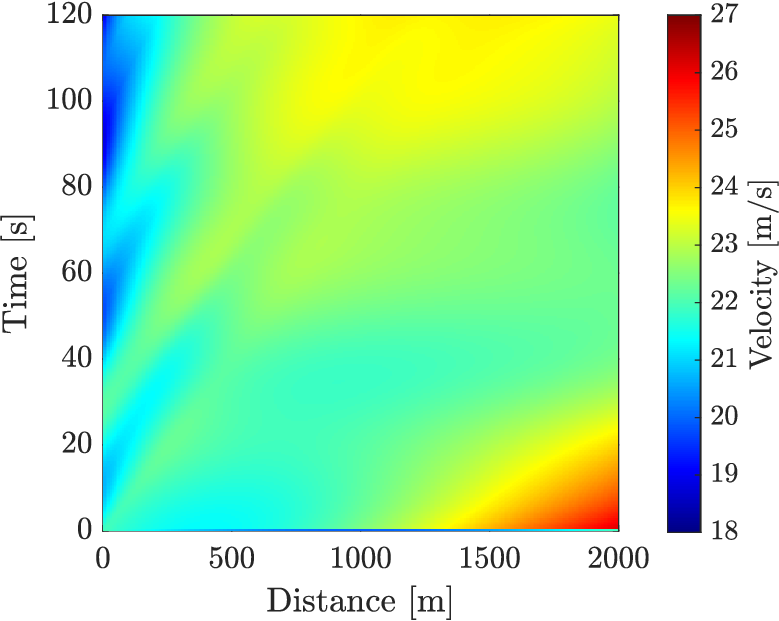} &
    \includegraphics[width=.6\columnwidth]{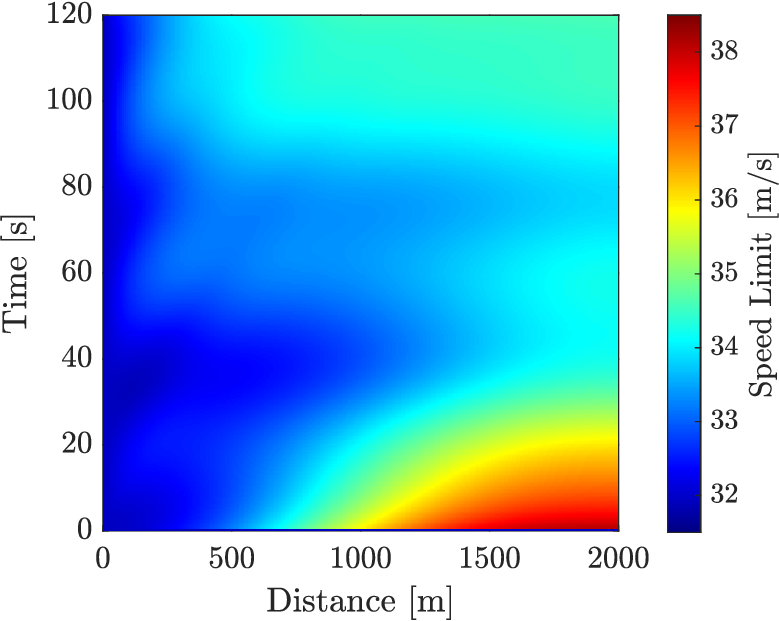} \\
    \small (a) &
    \small (b) &
    \small (c)
  \end{tabular}
  \caption{Density (a), velocity (b), and variable speed limit (c) of nonlinear model with $Q_0 = 5e-5$.}
  \label{fig:nonlin_cont_results}
\end{figure*}

\subsection{Comparison}
As shown in both Fig. \ref{fig:lin_totcars} and Fig. \ref{fig:nonlin_totcars}, the performance of the controller differs between the linear and nonlinear implementation. In each case, for the same value of $Q_0$, the linear model is affected more than the nonlinear model. For example, when $Q_0 = 5e-4$ the controller is able to drive the system to the desired traffic density in the linear model, but cannot do it for the nonlinear model. Similarly, the linear model reaches the desired density at 20 s, whereas in the nonlinear model it does not get close until 40 s.

To further investigate the performance of the designed controller on the two models Fig. \ref{fig:lin_cont_results} and Fig. \ref{fig:nonlin_cont_results} show the resulting density, velocity, and speed limit for the linear and nonlinear model, respectively. The results shown are for $Q_0 = 5e-5$. Looking at Fig. \ref{fig:lin_cont_results}(a) and Fig. \ref{fig:nonlin_cont_results}(a), it can be seen that the nonlinear controller is not as effective. The high density region resulting from the initial condition at $x = 500$ m to $x = 1500$ m is dissipated in the linear model, and it never reaches the downstream boundary. On the other hand, in the nonlinear model, the high density region continues to move until it reaches the downstream boundary at around 20 s. As well, the upstream boundary conditions propagate further into the road in the nonlinear simulation, but in the linear simulation the controller is able to mitigate them quickly.

Another difference can be seen in the speed limits for the linear and nonlinear model, Fig. \ref{fig:lin_cont_results}(c) and Fig. \ref{fig:nonlin_cont_results}(c), respectively. In the nonlinear simulation, the controller needs to keep a higher speed limit for longer at the end of the length of road. The high speed limit lasts until just before 20 s in the nonlinear model whereas in the linear model the high speed limit region ends after around 5 s.

As explained earlier, the control input for the linear model is the rate of change of the VSL rate over space $\frac{\partial b}{\partial z}$, but in the nonlinear model it is the actual VSL rate, $b$. This could explain the lag in response of the nonlinear model when compared to the linear model.

\section{CONCLUSIONS}
In this paper, a LQ controller was designed for the LWR model using variable speed limit control. First, the linearized LWR model was defined and a variable speed limit was applied to the model as a modification of Greenshield's model. Then, a state feedback function was computed via the solution of a Riccati differential equation. In this application, instead of the control input being just a distributed parameter, the control input was the rate of change of the distributed parameter. The designed controller was then verified on both the linear and nonlinear LWR model, and the performance of the two controllers was compared. Future work will investigate how to tune $Q_0$ as well as how to develop a controller that can be used in either a mixed traffic or congested traffic scenario to mitigate traffic jams in the presence of shocks.

\bibliographystyle{IEEEtran}
\bibliography{root}

\begin{thebibliography}{10}
\providecommand{\url}[1]{#1}
\csname url@rmstyle\endcsname
\providecommand{\newblock}{\relax}
\providecommand{\bibinfo}[2]{#2}
\providecommand\BIBentrySTDinterwordspacing{\spaceskip=0pt\relax}
\providecommand\BIBentryALTinterwordstretchfactor{4}
\providecommand\BIBentryALTinterwordspacing{\spaceskip=\fontdimen2\font plus
\BIBentryALTinterwordstretchfactor\fontdimen3\font minus
  \fontdimen4\font\relax}
\providecommand\BIBforeignlanguage[2]{{%
\expandafter\ifx\csname l@#1\endcsname\relax
\typeout{** WARNING: IEEEtran.bst: No hyphenation pattern has been}%
\typeout{** loaded for the language `#1'. Using the pattern for}%
\typeout{** the default language instead.}%
\else
\language=\csname l@#1\endcsname
\fi
#2}}

\bibitem{schrank2012umr}
\BIBentryALTinterwordspacing
D.~Schrank, T.~Lomax, and B.~Eisele, ``2021 urban mobility report,'' Texas
  Transportation Institute, 2021. [Online]. Available:
  \url{http://mobility.tamu.edu/ums/report}
\BIBentrySTDinterwordspacing

\bibitem{pishue2022inrix}
B.~Pishue, ``2022 {INRIX} {G}lobal {T}raffic {S}corecard,'' in \emph{INRIX
  (January 2023)}, 2023.

\bibitem{cookson2017inrix}
G.~Cookson and B.~Pishue, ``2016 {INRIX} {G}lobal {T}raffic {S}corecard,'' in
  \emph{INRIX (January 2017)}, 2017.

\bibitem{treiber2012traffic}
M.~Treiber and A.~Kesting, \emph{Traffic Flow Dynamics: Data, Models and
  Simulation}.\hskip 1em plus 0.5em minus 0.4em\relax Springer Science \&
  Business Media, 2012.

\bibitem{lighthill1955kinematic}
M.~J. Lighthill and G.~B. Whitham, ``On kinematic waves ii. a theory of traffic
  flow on long crowded roads,'' \emph{Proceedings of the Royal Society of
  London. Series A. Mathematical and Physical Sciences}, vol. 229, no. 1178,
  pp. 317--345, 1955.

\bibitem{richards1956shock}
P.~I. Richards, ``Shock waves on the highway,'' \emph{Operations research},
  vol.~4, no.~1, pp. 42--51, 1956.

\bibitem{tumash2019robust}
L.~Tumash, C.~Canudas-de Wit, and M.~L. Delle~Monache, ``Robust tracking
  control design for fluid traffic dynamics,'' in \emph{2019 IEEE 58th
  Conference on Decision and Control (CDC)}.\hskip 1em plus 0.5em minus
  0.4em\relax IEEE, 2019, pp. 4085--4090.

\bibitem{knoop2013traffic}
V.~Knoop, \emph{Introduction to Traffic Flow Theory: An introduction with
  exercises}.\hskip 1em plus 0.5em minus 0.4em\relax TU Delft Open, 2017.

\bibitem{carlson2011local}
R.~C. Carlson, I.~Papamichail, and M.~Papageorgiou, ``Local feedback-based
  mainstream traffic flow control on motorways using variable speed limits,''
  \emph{IEEE Transactions on intelligent transportation systems}, vol.~12,
  no.~4, pp. 1261--1276, 2011.

\bibitem{spiliopoulou2014macroscopic}
A.~Spiliopoulou, M.~Kontorinaki, M.~Papageorgiou, and P.~Kopelias,
  ``Macroscopic traffic flow model validation at congested freeway off-ramp
  areas,'' \emph{Transportation Research Part C: Emerging Technologies},
  vol.~41, pp. 18--29, 2014.

\bibitem{yu2020bilateral}
H.~Yu, M.~Diagne, L.~Zhang, and M.~Krstic, ``Bilateral boundary control of
  moving shockwave in {LWR} model of congested traffic,'' \emph{IEEE
  Transactions on Automatic Control}, vol.~66, no.~3, pp. 1429--1436, 2020.

\bibitem{liu2021robust}
H.~Liu, C.~Claudel, and R.~Machemehl, ``Robust traffic control using a first
  order macroscopic traffic flow model,'' \emph{IEEE Transactions on
  Intelligent Transportation Systems}, vol.~23, no.~7, pp. 8048--8062, 2021.

\bibitem{yu2018adaptive}
H.~Yu and M.~Krstic, ``Adaptive output feedback for aw-rascle-zhang traffic
  model in congested regime,'' in \emph{2018 Annual American Control Conference
  (ACC)}.\hskip 1em plus 0.5em minus 0.4em\relax IEEE, 2018, pp. 3281--3286.

\bibitem{yu2019traffic}
------, ``Traffic congestion control for {Aw-Rascle-Zhang} model,''
  \emph{Automatica}, vol. 100, pp. 38--51, 2019.

\bibitem{yu2019reinforcement}
H.~Yu, S.~Park, A.~Bayen, S.~Moura, and M.~Krstic, ``Reinforcement learning
  versus {PDE} backstepping and {PI} control for congested freeway traffic,''
  \emph{arXiv preprint arXiv:1904.12957}, 2019.

\bibitem{pasquale2018closed}
C.~Pasquale, S.~Sacone, and S.~Siri, ``Closed-loop stability of freeway traffic
  systems with ramp metering control,'' in \emph{2018 IEEE Conference on
  Decision and Control (CDC)}.\hskip 1em plus 0.5em minus 0.4em\relax IEEE,
  2018, pp. 223--228.

\bibitem{tumash2021boundary}
L.~Tumash, C.~C. de~Wit, and M.~L. Delle~Monache, ``Boundary control design for
  traffic with nonlinear dynamics,'' \emph{IEEE Transactions on Automatic
  Control}, 2021.

\bibitem{papamichail2008integrated}
I.~Papamichail, K.~Kampitaki, M.~Papageorgiou, and A.~Messmer, ``Integrated
  ramp metering and variable speed limit control of motorway traffic flow,''
  \emph{IFAC Proceedings Volumes}, vol.~41, no.~2, pp. 14\,084--14\,089, 2008.

\bibitem{liu2016model}
S.~Liu, H.~Hellendoorn, and B.~De~Schutter, ``Model predictive control for
  freeway networks based on multi-class traffic flow and emission models,''
  \emph{IEEE Transactions on Intelligent Transportation Systems}, vol.~18,
  no.~2, pp. 306--320, 2016.

\bibitem{karafyllis2019feedback}
I.~Karafyllis and M.~Papageorgiou, ``Feedback control of scalar conservation
  laws with application to density control in freeways by means of variable
  speed limits,'' \emph{Automatica}, vol. 105, pp. 228--236, 2019.

\bibitem{pisarski2015nash}
D.~Pisarski and C.~Canudas-de Wit, ``Nash game-based distributed control design
  for balancing traffic density over freeway networks,'' \emph{IEEE
  Transactions on Control of Network Systems}, vol.~3, no.~2, pp. 149--161,
  2015.

\bibitem{moghadam2010lq}
A.~A. Moghadam, I.~Aksikas, S.~Dubljevic, and J.~F. Forbes, ``Lq control of
  coupled hyperbolic pdes and odes: Application to a cstr-pfr system,''
  \emph{IFAC Proceedings Volumes}, vol.~43, no.~5, pp. 721--726, 2010.

\bibitem{aksikas2009lq}
I.~Aksikas, A.~Fuxman, J.~F. Forbes, and J.~J. Winkin, ``Lq control design of a
  class of hyperbolic pde systems: Application to fixed-bed reactor,''
  \emph{Automatica}, vol.~45, no.~6, pp. 1542--1548, 2009.

\bibitem{carlson2011mainstream}
R.~Carlson, ``Mainstream traffic flow control on motorways,'' Ph.D.
  dissertation, Ph. D. dissertation, Tech. Univ. Crete, Chania, Greece, 2011.

\end{thebibliography}

\end{document}